%% file: main.tex
\documentclass{article}

\pdfoutput=1 %

\usepackage[nonatbib,final]{neurips_data_2021}
\usepackage[round,semicolon]{natbib}

\usepackage[utf8]{inputenc} %
\usepackage[T1]{fontenc}    %
\usepackage{dsfont}

\usepackage[breaklinks=true,colorlinks,citecolor=black,bookmarks=false]{hyperref}
\hypersetup{
	colorlinks=true,
	linkcolor=blue,
	filecolor=magenta,      
	urlcolor=blue,
	citecolor=black,
  pdfinfo={
    Title={Measuring Coding Challenge Competence With APPS},
    Author={Dan Hendrycks and Steven Basart and Saurav Kadavath and Mantas Mazeika and Akul Arora and Ethan Guo and Collin Burns and Samir Puranik and Horace He and Dawn Song and Jacob Steinhardt},
    Subject={coding, transformers},
    Keywords={program synthesis, code, interview, gpt, reasoning, logic, problem solving, transformers, automated programming, apps dataset}
  }
}

\usepackage{url}            %
\usepackage{booktabs}       %
\usepackage{amsfonts}       %
\usepackage{nicefrac}       %
\usepackage{microtype}      %

\usepackage[frozencache=true,cachedir=.]{minted}
\usepackage{listings}
\usepackage{xcolor}

\definecolor{codepurple}{rgb}{0.58,0,0.82}

\definecolor{codegreen}{rgb}{0,0.5,0}
\definecolor{codeblue}{rgb}{0.25,0.5,0.5}
\definecolor{codeblue2}{rgb}{0.1,0.1,0.8}
\definecolor{codegray}{rgb}{0.6,0.6,0.6}

\definecolor{backcolour}{rgb}{0.97, 0.97, 0.97}
\definecolor{white}{rgb}{1, 1, 1}
\definecolor{backcolour2}{rgb}{0.90, 0.90, 0.90}

\lstdefinestyle{mystyle}{
    backgroundcolor=\color{white},   
    commentstyle=\color{codegray},
    keywordstyle=\color{codeblue2},
    numberstyle=\tiny\color{codegray},
    stringstyle=\color{codeblue},
    basicstyle=\fontsize{9pt}{9.5pt}\fontfamily{lmtt}\selectfont,
    breakatwhitespace=false,         
    breaklines=true,                 
    captionpos=b,                    
    keepspaces=true,                 
    numbers=left,                    
    numbersep=12pt,                  
    showspaces=false,                
    showstringspaces=false,
    showtabs=false,                  
    tabsize=2,
    framexleftmargin=20pt,
    framextopmargin=10pt,
    framesep=30pt,
}

\usepackage{times}
\usepackage{epsfig}
\usepackage{wrapfig}
\usepackage{lipsum}

\usepackage[utf8]{inputenc} %
\usepackage{amsmath, amssymb}
\usepackage[T1]{fontenc}    %
\usepackage{url}            %
\usepackage{amsfonts}       %
\usepackage{microtype}      %
\usepackage{wrapfig}
\usepackage{pdfpages}
\usepackage{tabularx}
\usepackage{arydshln}
\usepackage{multirow}
\usepackage{caption}
\usepackage{subcaption}
\usepackage{hhline}
\usepackage{multirow}
\usepackage{tabularx}
\newcolumntype{Y}{>{\centering\arraybackslash}X}
\usepackage{multirow}
\usepackage{makecell}
\usepackage{mathtools}
\usepackage[utf8]{inputenc}
\usepackage{stfloats}
\usepackage{spverbatim}
\usepackage{makecell}

\usepackage{cleveref}
\usepackage{appendix}

\usepackage{tikz}
\definecolor{rightgreen}{RGB}{0,154,24}
\def\checkmark{\tikz\fill[scale=0.4](0,.35) -- (.25,0) -- (1,.7) -- (.25,.15) -- cycle;} 
\usepackage{pifont}%
\newcommand{\xmark}{\ding{53}}%

\makeatletter
\newcommand*\bigcdot{\mathpalette\bigcdot@{.5}}
\newcommand*\bigcdot@[2]{\mathbin{\vcenter{\hbox{\scalebox{#2}{$\m@th#1\bullet$}}}}}
\makeatother

\makeatletter
\newcommand{\printfnsymbol}[1]{%
  \textsuperscript{\@fnsymbol{#1}}%
}
\makeatother

\newcommand{\reviewer}[3]{
	\expandafter\newcommand\csname #1\endcsname[1]{
		\textcolor{#3}{[#2: ##1]}
	}
}
\reviewer{norman}{Norman}{green}
\definecolor{neonpurple}{rgb}{0.3,0,1}
\reviewer{dan}{Dan}{neonpurple}
\reviewer{mantas}{Mantas}{red}
\reviewer{justin}{Justin}{blue}

\title{Measuring Coding Challenge Competence With APPS}
\date{}

\author{Dan Hendrycks\thanks{Equal Contribution.}\\
UC Berkeley\\
\And
Steven Basart\printfnsymbol{1}\\
UChicago\\
\And
Saurav Kadavath\\
UC Berkeley\\
\And
Mantas Mazeika\\
UIUC\\
\And
Akul Arora\\
UC Berkeley\\
\And
Ethan Guo\\
UC Berkeley\\
\And
Collin Burns\\
UC Berkeley\\
\And
Samir Puranik\\
UC Berkeley\\
\And
Horace He\\
Cornell\\
\AND
Dawn Song\\
UC Berkeley\\
\And
Jacob Steinhardt\\
UC Berkeley\\
}

\begin{document}
\maketitle

\begin{abstract}
While programming is one of the most broadly applicable skills in modern society, it is unclear how well state-of-the-art machine learning models can write code. Despite its importance, there has been surprisingly little work on evaluating code generation, and it can be difficult to assess code generation performance in an accurate and rigorous manner. To meet this challenge, we introduce APPS, a benchmark for code generation. Unlike prior work in more restricted settings, our benchmark measures the ability of models to take an arbitrary natural language specification and generate satisfactory Python code. Similar to how companies assess candidate software developers, we evaluate models by checking their generated code on test cases. Our benchmark includes $10,\!000$ problems, which range from having simple one-line solutions to being substantial algorithmic challenges. We fine-tune large language models on both GitHub and our training set, and we find that the prevalence of syntax errors is decreasing exponentially as models improve. Recent models such as GPT-Neo can pass approximately $20\%$ of the test cases of introductory problems, so we find that machine learning models are now beginning to learn how to code. As the social significance of automatic code generation increases over the coming years, our benchmark can provide an objective measure for tracking advancements.\looseness=-1
\end{abstract}

``Everybody should learn to program a computer, because it teaches you how to think.'' -- \emph{Steve Jobs}

\input{sections/1-intro}

\input{sections/2-related-work}

\input{sections/3-dataset}
\input{sections/4-experiments}
\input{sections/5-conclusion}

\bibliography{main}
\bibliographystyle{plainnat}

\newpage

\appendix

\input{sections/6-appendix}

\end{document}

%% file: sections/1-intro.tex
\section{Introduction}
Computer programming can be found in nearly all parts of society. Spanning entertainment, healthcare, education, and more, programming is an extraordinarily general tool with applications that are vast in scope. As computers are becoming more ubiquitous in modern life, rising demand for high-quality code draws an ever-greater number of aspiring programmers to the profession. After years of study to become proficient coders, human experts are are able to convert abstract specifications of diverse cognitive tasks into concrete programs.

In the past few years, large-scale language models have shown promise in generalizing to various cognitive tasks, including linguistic inference \citep{NEURIPS2019_4496bf24}, commonsense reasoning \citep{zellers2019hellaswag,huang2019cosmosqa,bisk2019physicaliqa}, logical deduction \citep{Liu2020LogiQAAC}, mathematics \citep{Polu2020GenerativeLM,hendrycksmath2021}, and general understanding of multiple domains of human knowledge \citep{hendryckstest2021}. However, whether large-scale language models can reliably write code remains an open question. %

Motivated by the potential of language models and the need for thorough code generation evaluation, we introduce APPS, a benchmark for code generation from natural language specifications. Unlike prior work on code generation with Transformer language models \citep{Vaswani2017AttentionIA}, which mostly focuses on code translation \citep{lachaux2020unsupervised} and pseudocode-to-code \citep{NEURIPS2019_7298332f}, we evaluate models on their ability to take specifications given in natural language and write code that meets these specifications. This setting mirrors how human coders are evaluated and is a more realistic and informative setting in which to benchmark models.

\begin{figure*}[t]
    \centering
    \includegraphics[width=\textwidth]{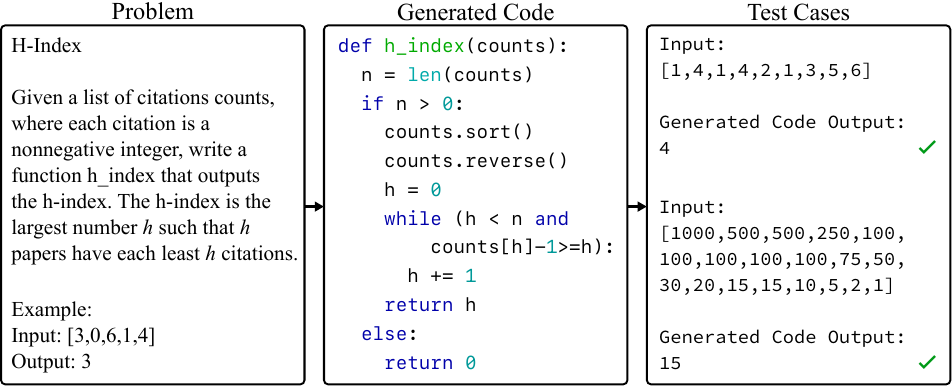}
    \caption{An example problem from APPS (left) along with possible generated code (middle) and two example test cases we use to evaluate the generated code (right). Our evaluation framework has test cases and $10,\!000$ code generation problems of varying difficulty levels.}
    \label{fig:apps_splash}
\end{figure*}

APPS provides a precise and comprehensive view of code generation. APPS evaluates models not only on their ability to code syntactically correct programs, but also on their ability to understand task descriptions and devise algorithms to solve these tasks. It contains $10,\!000$ programming problems at various levels of difficulty, covering simple introductory problems, interview-level problems, and coding competition challenges. If a model were to perform well on APPS, this would indicate an ability to flexibly use data structures and programming techniques, as well as an ability to correctly interpret diverse task specifications, follow instructions, and understand human intent \citep{hendrycks2021ethics}.

For most text generation tasks, high-quality evaluation requires human feedback, which can be time-consuming or carry pecuniary costs. As a result, automatic metrics such as BLEU \citep{papineni2002bleu} are often used to compare methods, but these metrics do not necessarily track program correctness. Since the objective for code generation is to produce correct programs, we assess programs not with BLEU but with test cases and error catching. Evaluating code generation on APPS is facilitated by a large bank of over $130,\!000$ test cases. The test cases are specifically chosen to probe correct functionality across the input space. By using test cases, we provide a gold-standard metric for code generation quality.

In our experiments, we find that models are now starting to exhibit nonzero accuracy and solve some coding problems. Additionally, as models improve, we observe that syntax errors are exponentially decreasing. We also find further evidence that BLEU is a problematic metric for code generation, sometimes being anticorrelated with gold-standard accuracy. We find that accuracy decreases with difficulty level and improves through fine-tuning and model size increases. The strongest model that we evaluate on introductory problems passes almost $20\%$ of test cases given five attempts. These results position code generation as a challenging but now tractable testbed for large-scale language models.

Writing code to meet specifications in natural language is an economically valuable task with widespread social implications should it be solved, as it could eventually facilitate malicious code generation and one day result in job automation. As large-scale language models have the potential to make significant progress on code generation, it is essential that we begin to track advancements on this task. Our new benchmark facilitates measuring performance in an accurate and rigorous manner. Using APPS, we find that programming is very difficult for modern language models, though performance is improving. Thus, the APPS benchmark can provide foresight about the performance of future large-scale language models at the critical task of program synthesis from natural language. The dataset is available at \href{https://github.com/hendrycks/apps}{https://github.com/hendrycks/apps}.

%% file: sections/2-related-work.tex
\begin{table*}[t]
\setlength{\tabcolsep}{9pt}
\fontsize{10}{11}\selectfont
\centering
\begin{tabular}{lcccc}
& PY150 & CONCODE & SPoC & APPS \\
\hline
Programming Language             & Python & Java & C++ & Python \\
Test Cases             & {\color{red}\xmark} & {\color{red}\xmark} & {\color{rightgreen}\checkmark} & {\color{rightgreen}\checkmark} \\
Number of Programs             & N/A & 104,000 & 18,356 & 232,421 \\
Lines per Program (Avg.)     & 1 & 26.3 & 14.7 & 18.0 \\
Number of Exercises             & 3,000 & 104,000 & 677 & 10,000 \\
Text Input      & Python & Docstrings & Pseudocode & Problem Descriptions \\
\hline
\end{tabular}
\caption{A comparison of the APPS dataset to existing datasets for converting between text and code. APPS has over an order of magnitude more ground-truth solutions than these datasets, test cases, and natural language problem descriptions.}
\label{tab:dataset_comparison}
\end{table*}

\section{Related Work}

\textbf{Program Synthesis.}\quad
Program synthesis is the task of generating a computer program that satisfies given specifications.
Deductive program synthesis uses formal logic specifications to define a search problem. Complex optimization techniques are used to generate programs satisfying these specifications \citep{alur2018sygus}. Because specifications must be converted into a formal language, these approaches can be rigid. Inductive synthesis from example input-output behavior can provide an alternative to formal specification \citep{Cai2017MakingNP,Gulwani2017ProgramS}, but it is often hard to full specify behavior with examples, as any machine learning practitioner is well-aware.

An alternative to formal or inductive specification is to specify program behavior in natural language, which prior work has considered in constrained settings.
\citet{Raza2015CompositionalPS} and \citet{desai2016program} generate short programs using ad-hoc programming languages to solve specifications such as ``Any 2 letters followed by any combination of 6 whole numbers.'' \citet{yu2018spider} introduce the Spider dataset for converting natural language queries into short SQL database commands. In contrast, we consider long natural language specifications and general-purpose programming languages.

\textbf{Code Understanding Datasets.}\quad
Language modeling is a compelling tool for code generation, and several works have achieved success generating code with language models in limited settings. \citet{lachaux2020unsupervised} use unsupervised machine translation techniques to translate functions across programming languages, attaining identical behavior after translation in many cases. \citet{NEURIPS2019_7298332f} introduce SPoC, a method for converting pseudocode to code utilizing seq2seq machine translation with an additional search step. To train SPoC, they collect line-by-line descriptions of C++ programs using Amazon Mechanical Turk. Recently, \citet{Lu2021CodeXGLUEAM} introduce the CodeXGLUE benchmark which aggregates various previous benchmarks and use CodeBLEU \citep{Ren2020CodeBLEUAM} and CONCODE. \citet{iyer-etal-2018-mapping} investigate generating Java code from docstrings and evaluate performance with BLEU. The docstrings are often incomplete specifications of what should be coded and only $14.7$ words long on average, e.g. ``Convert mixed case to underscores.'' By comparison, problem specifications in our new APPS benchmark are self-contained and have a much larger average length of $293.2$ words. Unlike \citet{iyer-etal-2018-mapping}, APPS contains test cases for every exercise, enabling a high-quality evaluation of code correctness. Further comparisons are in the Appendix.

\textbf{Evaluating Large-Scale Language Models.}\quad
Modern large-scale language models have demonstrated impressive capabilities across a variety of text-based tasks. On the SuperGLUE benchmark \citep{Wang2019SuperGLUEAS}, some models now exceed human performance. On many commonsense reasoning benchmarks, performance is rising quickly \citep{zellers2019hellaswag,huang2019cosmosqa,bisk2019physicaliqa}. Even when language models are evaluated across diverse technical areas such as law and medicine, performance is surprisingly high and poised to improve as models are scaled up further \citep{hendryckstest2021}. With rapid improvements across numerous datasets, finding resilient benchmarks on which models significantly underperform humans is challenging. APPS represents an attempt to fill this gap and cleanly separate model performance from that of expert humans.

%% file: sections/3-dataset.tex
\section{The APPS Dataset}\label{sec:dataset}

The APPS dataset consists of problems collected from different open-access coding websites such as Codeforces, Kattis, and more. The APPS benchmark attempts to mirror how humans programmers are evaluated by posing coding problems in unrestricted natural language and using test cases to evaluate solution correctness. The problems range in difficulty from introductory to collegiate competition level and measure coding and problem-solving ability. 

The Automated Programming Progress Standard, abbreviated APPS, consists of $10,\!000$ coding problems in total, with $131,\!777$ test cases for checking solutions and $232,\!421$ ground-truth solutions written by humans. Problems can be complicated, as the average length of a problem is $293.2$ words.
The data are split evenly into training and test sets, with $5,\!000$ problems each. In the test set, every problem has multiple test cases, and the average number of test cases is $21.2$. Each test case is specifically designed for the corresponding problem, enabling us to rigorously evaluate program functionality.

\noindent\textbf{Dataset Construction.}\quad
To create the APPS dataset, we manually curate problems from open-access sites where programmers share problems with each other, including Codewars, AtCoder, Kattis, and Codeforces. Problems are posed as natural language specifications of what should be coded, and they come in various formats. To improve quality and consistency, we wrote custom HTML parsers for each source of problems, which allows us to properly format LaTeX expressions, lists, and sections in the question text. Where necessary, we convert equation images to LaTeX using the MathPix API, and we remove problems that rely on image figures. We also perform deduplication using tf-idf features with SVD dimensionality reduction and cosine similarity. Several graduate and undergraduate student authors polished and refined this dataset over the course of six months, ensuring a high-quality set of problems.

Executing and evaluating arbitrary Python code is challenging. On the websites we source data from, human solutions are allowed to run arbitrary code, including import statements for common modules and libraries. To handle this, each website implements a custom judging system for solutions. We design a testing framework with this in mind, which merges the judging functionality of several websites. We also standardize the format of test cases. The end result is that solutions are allowed to execute arbitrary Python code, and the results are compared against test cases for a given problem.

\noindent\textbf{Dataset Difficulty.}\quad
Each of our problem sources uses a separate scale for measuring difficulty. We place problems from these different sources into three categories. For example, problems from Kattis with difficulty less than $3$ are categorized as ``introductory,'' problems with difficulty between $3$ and $5$ as ``interview,'' and problems with difficulty greater than $5$ as ``competition.''

\begin{enumerate}
    \item \textbf{Introductory Level}. These are problems that most programmers with 1-2 years of experience can answer without requiring complicated algorithms. Examples of such problems include counting the number of vowels in a string, or returning the running sum of a list of integers. There are $3,\!639$ problems classified as introductory level and $1,\!000$ in the test set.
    \item \textbf{Interview Level}. These are problems that are more algorithmic and difficult in nature and would be at the level of questions asked in programming technical interviews. Examples of such problems might include those involving data structures such as trees or graphs, or problems that requiring nontrivial algorithms. There are $5,\!000$ problems classified as interview level and $3,\!000$ in the test set.
    \item \textbf{Competition Level}. These are problems are the most challenging and are at the level of the most advanced high school and collegiate programming competitions, including USACO, IOI, and ACM. There are $1,\!361$ competition level problems and $1,\!000$ in the test set.
\end{enumerate}

\noindent\textbf{Problem Formats.}\quad 
To accommodate a broad range of problem sources, problems in APPS come in two formats.
\begin{itemize}
    \item \textbf{Call-Based Format} problems generally provide initial starter code, usually in the form of a function header, and ask for the solution to be provided as the function's return value.
    \item \textbf{Standard Input Format} problems generally lack starter code. Instead, the model is only provided with the problem and must output its answers to the STDOUT stream, such as by using print statements.
\end{itemize}

For the call-based format problems, we prompt models using the following inputs:

\begin{spverbatim}
"\nQUESTION:\n" + q_str + "\n" + starter_code_str + "\n" + "\nUse Call-Based Format\n\nANSWER:\n"
\end{spverbatim}

For the above prompt, the variable \verb|q_str| represents the raw text of the problem statement. The variable \verb|starter_code_str| represents the starter code given in the problem definition, or the empty string if no starter code was provided. For the standard input format problems, we prompt the model with the input string as before, but we replace ``Call-Based Format'' with ``Standard Input Format.'' Note that if starter code is given, it is only part of the input. This means that to use the starter code, a model must learn to copy the starter code at the beginning of its outputted answer in order to get the question correct. We find that fine-tuned models are able to do this without difficulty.

\begin{figure*}
    \textbf{Problem} \\
     You are given a string $s=s_1s_2\dots s_n$ of length $n$, which only contains digits $1$, $2$,..., $9$. A substring $s[l \dots r]$ of $s$ is a string $s_l s_{l + 1} s_{l + 2} \ldots s_r$. A substring $s[l \dots r]$ of $s$ is called even if the number represented by it is even. Find the number of even substrings of $s$. Note, that even if some substrings are equal as strings, but have different $l$ and $r$, they are counted as different substrings. The first line contains an integer $n$ ($1 \le n \le 65000$) — the length of the string $s$. The second line contains a string $s$ of length $n$. The string $s$ consists only of digits $1$, $2$,..., $9$. Print the number of even substrings of $s$.\looseness=-1

 \textbf{Model Output}
 \begin{minted}{python}
n = int(input())
s = input()
ans = 0
for i in range(n):
    for j in range(i, n):
        if int(s[i:j+1]) %
            ans += 1
print(ans)
 \end{minted}

\caption{Model-generated code from GPT-2 1.5B. 18/18 test cases pass for the problem. Note that models were given in problems exactly as they they are formatted here, which even meant parsing \LaTeX. For brevity, we exclude formatting instructions from this depicted problem.}
    \label{fig:samples_from_1500}
\end{figure*}

\begin{figure*}
    \textbf{Problem} \\
     You are given two integers $n$ and $m$. Calculate the number of pairs of arrays $(a, b)$ such that:  the length of both arrays is equal to $m$;  each element of each array is an integer between $1$ and $n$ (inclusive);  $a_i \le b_i$ for any index $i$ from $1$ to $m$;  array $a$ is sorted in non-descending order;  array $b$ is sorted in non-ascending order. As the result can be very large, you should print it modulo $10^9+7$. Input: The only line contains two integers $n$ and $m$ ($1 \le n \le 1000$, $1 \le m \le 10$). Output: Print one integer – the number of arrays $a$ and $b$ satisfying the conditions described above modulo $10^9+7$.
 \\
 \\
 \textbf{Model Output}
 \begin{small}
 \begin{minted}{python}
n, m = map(int, input().split())
mod = 10**9+7

fact = [1, 1]
factinv = [1, 1]
inv = [0, 1]

for i in range(2, n+1):
    fact.append((fact[-1] * i) %
    inv.append((-inv[mod %
    factinv.append((factinv[-1] * inv[-1]) %

def cmb(n, r, mod):
    if (r < 0) or (n < r):
        return 0
    r = min(r, n - r)
    return fact[n] * factinv[r] * factinv[n-r] %

print((cmb(n, m, mod) * cmb(m-1, n-1, mod)) %
\end{minted}
 \end{small}
    \caption{An example from GPT-2 1.5B. Although the code generated passes 0 test cases, it looks plausible at first glance.}
    \label{fig:interesting_sample_from_1500}
\end{figure*}

\paragraph{Test Case Quality.}
In the APPS test split, the average number of test cases is $21.2$, but some problems only have two test cases. These problems mainly come from Kattis and were chosen for the test split due to limited numbers of competition problems. A potential concern is that these problems could result in false positives if models happen to guess both test cases correctly. This is very unlikely in problems with large output spaces, but some problems have small output spaces, such as $\{\text{``YES''}, \text{``NO''}\}$. Although the model must narrow down these two options from the space of all possible strings, we find that this is not difficult for current models to accomplish, and this may cause false positives in our evaluation.

To quantify the impact of these problems, we took all 2- and 3-test-case problems that GPT-Neo 2.7B passed and manually checked for false positives. Out of 12 problems that the model got correct with 2 test cases, 8 were false positives. Out of 6 problems with 3 test cases, only 1 was a false positive. Since the test cases are from coding challenge websites with many users, we can assume that false negatives are absent. Thus, the false positive rate with 2 test cases is $8/(890-4) = 0.009 < 1\%$ and $1/(765-5) = 0.001 < 1\%$ with 3 test cases. The upshot is that the potential for noise is significantly less than that of most naturally annotated datasets.

%% file: sections/4-experiments.tex
\section{Experiments}

\subsection{Experimental Setup}

\paragraph{Models.}
We use GPT-2 \citep{radford2019language}, GPT-3 \citep{Brown2020LanguageMA}, and GPT-Neo \citep{gpt-neo} models. The GPT architecture is especially appropriate for text generation because it is autoregressive. However, GPT-2 was not pretrained on code, so we pretrain it on GitHub as described in the next paragraph. Anecdotal evidence indicates that GPT-3 can generate code. To determine the extent of its code generation ability, we use the `davinci' (Instruct series) model, the largest publicly available model speculated to have 175 billion parameters. Finally, GPT-Neo has an architecture similar to GPT-3, and it was pretrained on the Pile \citep{gao2020pile} which includes GitHub. Unlike GPT-3, GPT-Neo's weights are publicly available, hence we are able to fine-tune it with APPS.

\paragraph{GPT-2 Pretraining.}
Since GPT-2 was trained on natural language and not code, we collected GitHub code to further pretrain GPT-2. GitHub repositories with fewer than one star were filtered out.
While Neo's GitHub pretraining data did \emph{not} undergo an APPS data decontamination process, our GPT-2 models are trained on decontaminated data.
Specifically, all repositories matching certain keywords that would suggest overlap with common programming exercises were removed. We provide the list of keywords in the Supplementary Materials. We also discard any GitHub code that contains functions with the same signatures as functions in the starter code in many of our APPS problems. This leaves us with 30 GB of Python code. To improve the efficiency of pretraining, we process all Python code in the pretraining dataset by converting from spaces to tabs, which saves the character conversion when running model tokenizers.

\paragraph{Fine-tuning.}
During fine-tuning with APPS, the objective is to predict the entire code solution, given both the English text problem statement and the problem format (call-based format or standard input format). For problems with starter code, we exclude the starter code from the training loss.

Across pretraining and fine-tuning, we use the AdamW optimizer \citep{Loshchilov2019DecoupledWD}, a batch size of $256$, and a weight decay of $0.05$. We fine-tune for $10$ epochs. We use DeepSpeed and its implementation of the ZeRO optimizer to reduce memory consumption while training large models \citep{Rasley2020DeepSpeedSO, rajbhandari2020zero}. Unless otherwise specified, we use the default HuggingFace generation parameters, except that we use beam search with a beam size of $5$. Models are fine-tuned on 8 A100 GPUs.

\subsection{Metrics}
To obtain a comprehensive evaluation of code generation ability, we use the large bank of test cases and ground-truth solutions provided with APPS. Test cases allow for \emph{automatic} evaluation, even though the the space of possible programs can be combinatorially large. Therefore, unlike many other text generation tasks, manual analysis is not necessary. We aggregate the generated code's performance on test cases with two metrics, ``test case average'' and ``strict accuracy.'' %

\textbf{Test Case Average.}\quad We compute the average fraction of test cases passed. 
Concretely, let the number of problems in the test set be $P$. For a given problem $p$, let the code generated to solve problem $p$ be denoted $\langle\texttt{code}_p\rangle$, and set of test cases for problem $p$ be $\{(x_{p,c},\, y_{p,c})\}_{c=1}^{C_p}$. Then the test case average is 
\[
\frac{1}{P}\sum_{p=1}^P \frac{1}{C_p} \sum_{c=1}^{C_p} \mathds{1}\{\texttt{eval}(\langle\texttt{code}_p\rangle,x_{p,c}) = y_{p,c}\}.
\]

Oftentimes, solutions can successfully pass a subset of the test cases but not cover every corner case. This allows for less stringent model evaluation, as strict accuracy may currently obscure model improvements.

\begin{table*}[t]
\setlength{\tabcolsep}{2pt}
\small
\centering
\begin{tabular}{lcccc|cccc}
\multicolumn{1}{l}{} &  \multicolumn{4}{c}{Test Case Average} & \multicolumn{4}{c}{Strict Accuracy} \\
Model       & Introductory & Interview & Competitive &  Average & Introductory & Interview & Competition &  Average \\

\toprule
GPT-2 0.1B          & 5.64 & 6.93 & 4.37 & 6.16 & 1.00 & 0.33 & 0.00 & 0.40 \\
GPT-2 1.5B          & 7.40 & 9.11 & 5.05 & 7.96 & 1.30 & 0.70 & 0.00 & 0.68 \\
GPT-Neo 2.7B        & 14.68 & 9.85 & 6.54 & 10.15 & 3.90 & 0.57 & 0.00 & 1.12 \\
GPT-3 175B        & 0.57 & 0.65 & 0.21 & 0.55 & 0.20 & 0.03 & 0.00 & 0.06 \\
\end{tabular}
\caption{Average percentage of test cases passed and strict accuracy for each model and difficulty level. All values are percentages. Note `0.1B' indicates the number of model parameters in billions. GPT-3 is a \emph{few-shot} model and not fine-tuned, unlike the other models. GPT-Neo does best and attains approximately 4\% strict accuracy on Introductory problems, and for these problems it passes approximately 15\% of the test cases.}
\label{tab:results}
\end{table*}

\textbf{Strict Accuracy.}\quad Eventually, generated solutions should pass all test cases including corner cases.
To compute the strict accuracy which requires programs pass every test case, we run the code generated by the model on every test case of every problem. Strict accuracy is then computed by taking the number of solutions passing every test case divided by the total number of exercises. Using the notation from before, we can write the strict accuracy as $\frac{1}{P}\sum_{p=1}^P \prod_{c=1}^{C_p} \mathds{1}\{\texttt{eval}(\langle\texttt{code}_p\rangle,x_{p,c}) = y_{p,c}\}.$ Future research may only use strict accuracy when models become sufficiently capable.

\subsection{Model Performance Analysis}

\paragraph{Qualitative Output Analysis.} Models can sometimes generate correct or superficially plausible code. \Cref{fig:samples_from_1500} shows code generated by GPT-2 1.5B that passes all test cases. When models do not pass the test cases, sometimes their generated code still appears plausible at first glance. For example, in \Cref{fig:interesting_sample_from_1500}, we see that the 1.5B parameter model generates code that is related to the problem statement and makes a plausible attempt to solve it.

\begin{figure}[t]
\begin{minipage}{.49\textwidth}
\centering
\includegraphics[width=\textwidth]{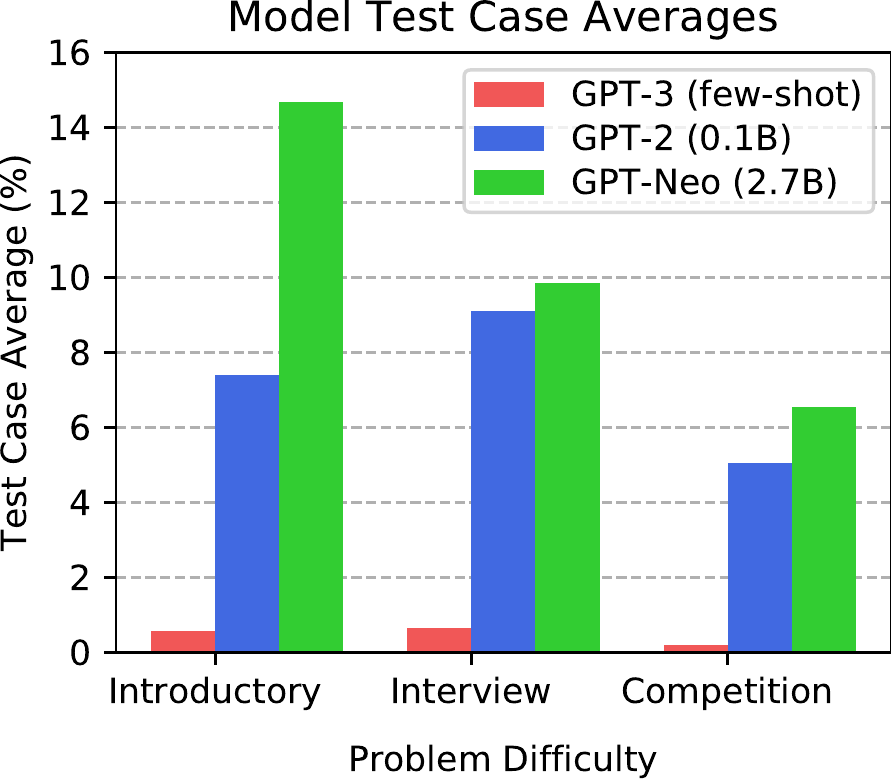}
\caption{The average percentage of test cases passed increases with larger fine-tuned models.%
}\label{fig:partial}
\end{minipage}\hfill%
\begin{minipage}{.49\textwidth}
\centering
\includegraphics[width=\textwidth]{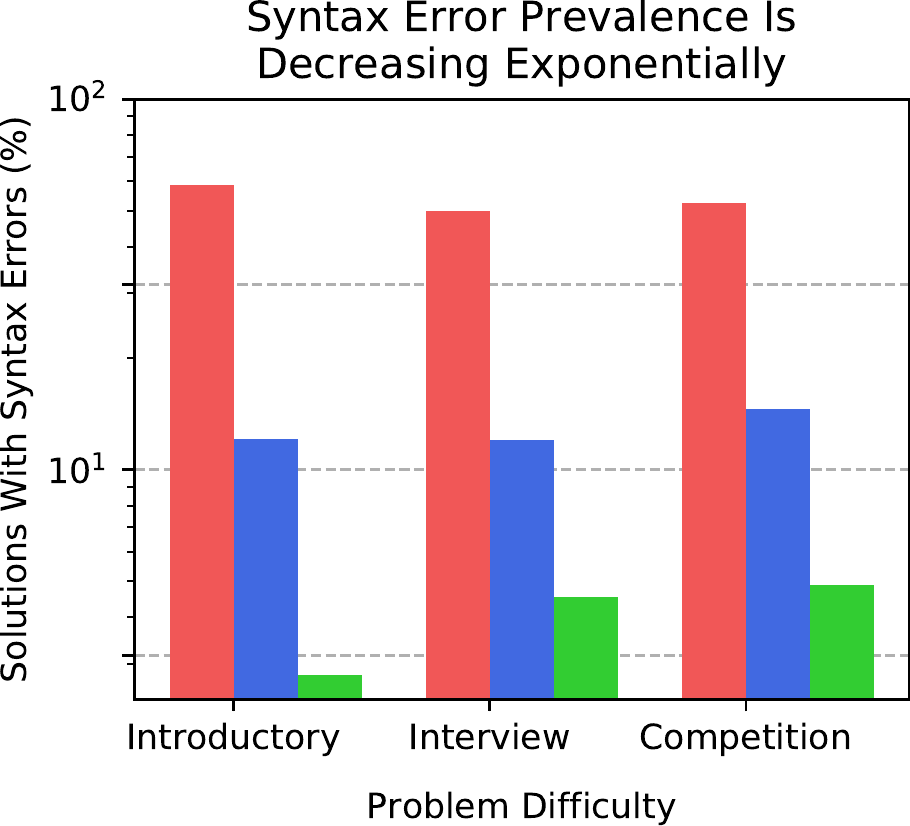}
\caption{Syntax errors decrease exponentially with fine-tuning and increased model sizes. GPT-Neo 2.7B has very few syntax errors.}\label{fig:syntax}
\end{minipage}
\end{figure}

\paragraph{Test Case Evaluation.}
We show the main results in \Cref{tab:results}. We observe that models are able to generate code that passed some test cases, implying many generated programs are free of syntax errors and can successfully process inputs test cases to produce correct answers. Note that for Introductory questions, GPT-Neo passes approximately $15\%$ of the test cases. We visualize Test Case Average results in \Cref{fig:partial}. This demonstrates models are showing marked improvements on code generation and now starting to have traction on code generation.

\begin{wraptable}{r}{0.43\textwidth}
\begin{center}
\begin{tabular}{l|cc}
    & Top-1 & Top-5 \\ \midrule
Test Case Average & $14.7\%$ & $19.9\%$ \\
Strict Accuracy & $3.9\%$ & $5.5\%$ \\
\bottomrule
\end{tabular}
\caption{GPT-Neo 2.7B performance on introductory problems using one generated program (Top-1) and the best of five generated programs (Top-5). Full results are in the Supplementary Materials. %
}
\label{tab:topk}
\end{center}
\end{wraptable}

Performance can be further improved by sampling multiple solutions and selecting the best. Here, we perform beam search with beam width $5$ and evaluate its $5$ beams, so that each model has five attempts to get a problem correct rather than one. With this setup, GPT-Neo's strict accuracy on Introductory problem then exceeds $5\%$, as shown in \Cref{tab:topk}. Our results in the Supplementary Materials show that the top-5 test case average GPT-2 0.1B is 10.75 while the top-1 test case average of GPT-2 1.5B is 7.96. This highlights that simply sampling multiple candidate solutions is a powerful way to markedly improve performance.

Our results also provide us with information about the importance of model choice. Evidently existing few-shot GPT-3 models are not necessarily better at code generation than fine-tuned models that are smaller by two orders of magnitude. Additionally, performance improvement from GPT-2 1.5B to GPT-Neo 2.7B is larger than that from GPT-2 0.1B to GPT-2 1.5B. Potential causes of GPT-Neo's better performance are that GPT-Neo is trained on more code from GitHub, it has more parameters, or its architecture hyperparameters were chosen better. Memorization explaining all performance is an implausible explanation as performance tracks problem difficulty; were models just memorizing, we would expect uniform performance across difficulties. Since models still have large room for improvement, solving the APPS benchmark without unreasonable amounts of computational resources may require architectural or algorithmic improvements.

\paragraph{Syntax Errors.} We now assess the frequency of syntax errors, errors that prevent the program from being interpreted including inconsistent spacing, unbalanced brackets, missing colons, and so on. Syntax errors are identified in our testing framework based on the heuristic of whether pyext is able to load the generated code as a Python module. For our purposes, this almost exclusively occurs for syntax errors. We visualize the prevalence of syntax errors in \Cref{fig:syntax}. While approximately $59\%$ of GPT-3's generated solutions for introductory problems have syntax errors, GPT-Neo syntax error frequency is approximately $3\%$. Note that recent work such as \cite{Yasunaga2020GraphbasedSP} create a separate model to repair source code to fix compilation issues, but our results suggest that such efforts may be unnecessary in the future as syntax error frequency is sharply decreasing automatically.

\begin{wrapfigure}{r}{0.5\textwidth}
    \centering
    \includegraphics[width=0.5\textwidth]{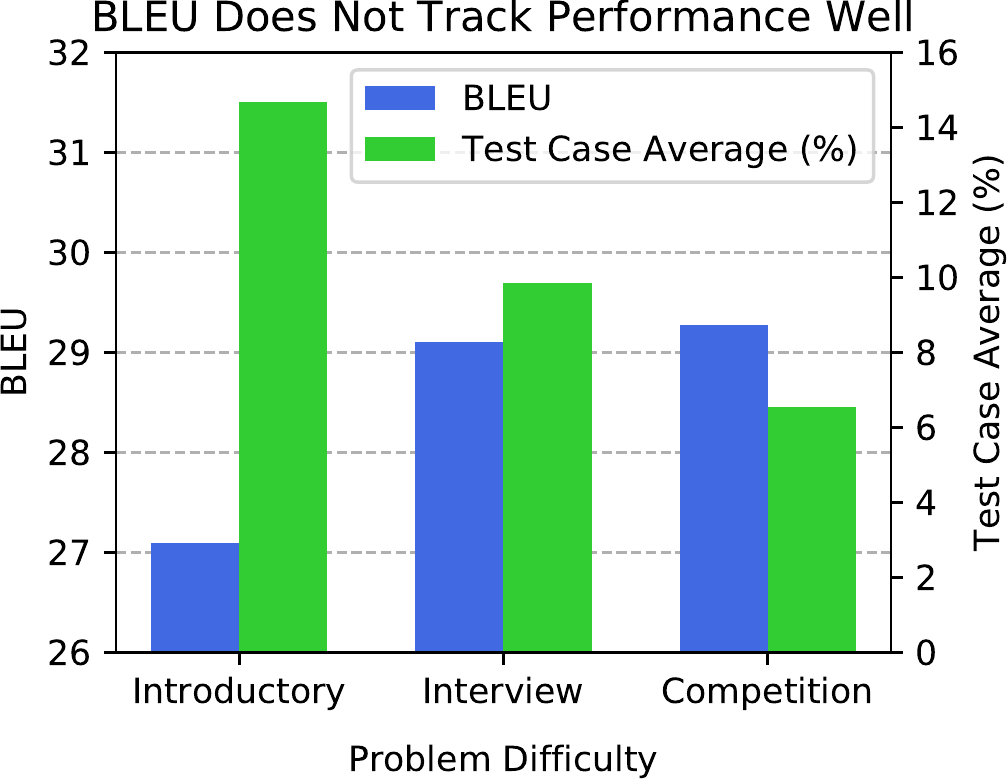}
    \caption{BLEU scores for GPT-Neo 2.7B increase with difficulty level and are anticorrelated with a gold-standard accuracy metric.}
    \label{fig:bleu}
\end{wrapfigure}

\paragraph{BLEU.} We find that assessing model performance with BLEU is a poor substitute for evaluating with test cases. To evaluate BLEU, we take the generated solution and compute its BLEU with each human-written solution for a given problem; we then record the highest BLEU score. Observe in \Cref{fig:bleu} that BLEU increases as problem sources become more difficult, even though models actually perform worse on harder problems. Moreover, worse models can have similar or higher BLEU scores. For example, GPT-2 0.1B has $26.8$, $29.7$, and $30.2$ as BLEU scores for introductory, interview, and competition problems, respectively. Meanwhile GPT-Neo 2.7B has $27.1$, $29.1$, and $29.3$ as its BLEU scores, respectively. Hence BLEU wrongly suggests GPT-Neo is a worse model.

\paragraph{Evaluating GPT-3.}
We evaluate GPT-3 175B on APPS in a few-shot setting. A separate prompt is used for standard input and call-based questions, and each prompt includes instruction text along with two example questions and solutions from the corresponding question type. We find that GPT-3 only solves $3$ problems out of $5,\!000$: two introductory problems and one interview problem. The two introductory problems are simple interpretation tasks, such as implementing a specified algebraic expression. The interview problem requires higher-level thinking that suggests nontrivial reasoning. However, it is possible that GPT-3 memorized the solution during pretraining, or that it took a lucky guess based on heuristics in the question. One potential factor in GPT-3's poor performance is that it handles syntax poorly. Namely, we observed cases where improper formatting of otherwise functioning code causes a syntax error. For specific examples and more details, see the Supplementary Materials.

\paragraph{Evaluations on Larger Models.}
Since the public release of APPS, several others have trained even larger models on APPS than we evaluate here. OpenAI Codex is a 12B parameter Transformer language model pre-trained on large quantities of public code and comments. \citet{chen2021evaluating} evaluate Codex on APPS under various configurations and achieve top-1 and top-5 accuracy on introductory problems of 4.14\% and 9.65\% respectively, close to double the top-5 accuracy of GPT-Neo 2.7B. Furthermore, by scaling up to a top-1000 evaluation they obtain 25\% accuracy. This demonstrates that larger models trained specifically for code generation can improve APPS performance even further, but are still far from solving the task.

%% file: sections/5-conclusion.tex
\section{Conclusion}
We introduced APPS, a benchmark of $10,\!000$ Python programming problems. Unlike prior work that focused on pseudocode to code generation or translation between programming languages, our benchmark measures how well language models can generate python code given natural language specifications. By performing extensive quality assurance and including hundreds of thousands of test cases and ground-truth solutions across different difficulty levels, we created a comprehensive and rigorous testbed for evaluating models. We assessed state-of-the-art generative models on our benchmark and found that overall performance was low. However, the prevalence of syntax errors decreased exponentially as models improved, and recent models such as GPT-Neo solved over $5\%$ of our introductory problems. As models become more competent at code generation, it is important to have a proxy for tracking this capability which could one day result in automation or malicious code generation. The APPS benchmark can provide an important measure for tracking upstream program synthesis advancements.

%% file: sections/6-appendix.tex
\begin{table*}[h]
\setlength{\tabcolsep}{9pt}
\fontsize{10}{11}\selectfont
\centering
\begin{tabular}{lccccc}
& Hearthstone & Django & NAPS & APPS \\
\hline
Programming Language             & Python & Python & UAST & Python \\
Test Cases             & {\color{red}\xmark} & {\color{red}\xmark} & {\color{rightgreen}\checkmark} & {\color{rightgreen}\checkmark} \\
Number of Programs             & 665 & 18,805 & 17,477 & 232,421 \\
Lines per Program (Avg.)     & 7.7 & 1 & 21.7 & 18.0 \\
Number of Exercises             & 665 & 18,805 & 2,231 & 10,000 \\
Text Input      & Card Text & Comment & Pseudocode & Problem Descriptions \\
\hline
\end{tabular}
\caption{Further comparisons of APPS with previous datasets.}
\label{tab:further_dataset_comparison}
\end{table*}

\begin{table*}[h]
\setlength{\tabcolsep}{2pt}
\small
\centering
\begin{tabular}{lcccc|cccc}
\multicolumn{1}{l}{} &  \multicolumn{4}{c}{Top-5 Test Case Average} & \multicolumn{4}{c}{Top-5 Strict Accuracy} \\
Model       & Introductory & Interview & Competitive &  Average & Introductory & Interview & Competition &  Average \\

\toprule
GPT-2 0.1B          & 13.81 & 10.97 & 7.03 & 10.75 & 2.70 & 0.73 & 0.00 & 1.02 \\
GPT-2 1.5B          & 16.86 & 13.84 & 9.01 & 13.48 & 3.60 & 1.03 & 0.00 & 1.34 \\
GPT-Neo 2.7B        & 19.89 & 13.19 & 9.90 & 13.87 & 5.50 & 0.80 & 0.00 & 1.58 \\

\end{tabular}
\caption{Top-5 performance of GPT-2 models and GPT-Neo. Taking the best of five candidate solutions markedly improves performance.}
\label{tab:top5full}
\end{table*}

\section{Auxiliary Dataset Information}\label{appendix:checklist}
\paragraph{Legal Compliance.}
In APPS, we scrape question text, ground-truth solutions, and test cases from various coding challenge websites. These websites are AtCoder, CodeChef, Codeforces, Codewars, HackerRank, Kattis, and LeetCode. In all cases, we only scrape public-facing data. For instance, we avoid scraping data from paywalled portions of sites. In the case of Kattis, all problems we scrape are under the CC BY-SA 3.0 license (\href{https://creativecommons.org/licenses/by-sa/3.0/}{https://creativecommons.org/licenses/by-sa/3.0/}). For other websites, some content may be copyrighted. In these cases, we abide by Fair Use §107: ``the fair use of a copyrighted work, including such use by ... scholarship, or research, is not an infringement of copyright'', where fair use is determined by ``the purpose and character of the use, including whether such use is of a commercial nature or is for nonprofit educational purposes'', ``the amount and substantiality of the portion used in relation to the copyrighted work as a whole'', and ``the effect of the use upon the potential market for or value of the copyrighted work.'' The APPS dataset is noncommercial and is likely to have no effect on the value of the original problems. Moreover, for all problem sources, we only scrape a fraction of the available problems and ground-truth solutions.

Regarding international copyright laws, the websites that we scrape from are based in the United States, Japan, India, and Russia, all of which are contracting parties to the WIPO Copyright Treaty. In the United States, the WIPO Copyright Treaty is implemented by the Digital Millenium Copyright Act (DMCA). Since APPS was made in the United States, the DMCA is the relevant legislation that we must comply with. Notably, DMCA §1201 states, ``No person shall circumvent a technological measure that effectively controls access to a work protected under this title.'' We do not circumvent access controls when creating APPS and hence abide by §1201. Fair Use extends to content protected by the DMCA, for which we refer readers to the previous paragraph.

Although GDPR only applies in the European Union, some of the ground-truth solutions in APPS may have been written by EU citizens. GDPR is chiefly concerned with the protection of personal data gathered by entities engaging in economic activity. The only personally linked information in APPS is the problem solutions written by individuals and published under aliases to public websites. In some cases, these solutions contain identifying information in comments, which we remove to preserve privacy. We comply with GDPR, because our processed solutions remove identifiers, and we are compliant because we collect the data for academic research purposes.

\paragraph{Author Statement and License.}
We bear all responsibility in case of violation of rights. The APPS data is licensed under CC BY-SA 3.0 in accordance with the Kattis problem licenses and the ShareAlike terms. Our code is open sourced under the MIT license.

\section{Datasheets}
We follow the recommendations of \citet{gebru2018datasheets} and provide a datasheet for the ETHICS dataset in this section.

\subsection{Motivation}

\paragraph{For what purpose was the dataset created? Was there a specific task
in mind? Was there a specific gap that needed to be filled? Please provide
a description.} 
The APPS dataset was created to track the progress of code generation models on the task of generating arbitrary Python code from complex natural language specifications, a challenging setting that had no rigorous benchmark before our work.

\paragraph{Who created the dataset (e.g., which team, research group) and on
behalf of which entity (e.g., company, institution, organization)?}
Refer to the main document.

\paragraph{Who funded the creation of the dataset? If there is an associated
grant, please provide the name of the grantor and the grant name and
number.}
There is no associated grant.

\paragraph{Any other comments?}
No.

\subsection{Composition}
\paragraph{What do the instances that comprise the dataset represent (e.g.,
documents, photos, people, countries)? Are there multiple types of
instances (e.g., movies, users, and ratings; people and interactions between them; nodes and edges)? Please provide a description.}
The instances are coding challenge problems posed in natural language, each of which consists of question text, ground-truth solutions, and test cases. Please refer to the main document for more detail.

\paragraph{How many instances are there in total (of each type, if appropriate)?}
APPS contains 10,000 problems, 232,421 ground-truth solutions, and 131,777 test cases.

\paragraph{Does the dataset contain all possible instances or is it a sample
(not necessarily random) of instances from a larger set? If the
dataset is a sample, then what is the larger set? Is the sample representative of the larger set (e.g., geographic coverage)? If so, please describe how
this representativeness was validated/verified. If it is not representative
of the larger set, please describe why not (e.g., to cover a more diverse
range of instances, because instances were withheld or unavailable).}
APPS contains a subset of all possible test cases for its problems. These test cases are written by problem designers to cover important functionality.

\paragraph{What data does each instance consist of? “Raw” data (e.g., unprocessed text or images) or features? In either case, please provide a description.}
Each instance consists of text and numerical data.

\paragraph{Is there a label or target associated with each instance? If so, please
provide a description.}
Each instance is associated with test cases, which provide a ground-truth signal for functional correctness.

\paragraph{Is any information missing from individual instances? If so, please
provide a description, explaining why this information is missing (e.g.,
because it was unavailable). This does not include intentionally removed
information, but might include, e.g., redacted text.}
No.

\paragraph{Are relationships between individual instances made explicit
(e.g., users’ movie ratings, social network links)? If so, please describe how these relationships are made explicit.}
We remove duplicate or near-duplicate problems from APPS.

\paragraph{Are there recommended data splits (e.g., training, development/validation,
testing)? If so, please provide a description of these splits, explaining
the rationale behind them.}
We provide a training and test split. The splits were optimized for increasing the number of test cases in the test split while maintaining a fixed number of problems from each difficulty.

\paragraph{Are there any errors, sources of noise, or redundancies in the
dataset? If so, please provide a description.}
See Section 3 in the main paper for a discussion of test case quality.

\paragraph{Is the dataset self-contained, or does it link to or otherwise rely on
external resources (e.g., websites, tweets, other datasets)?}
The dataset is self-contained.

\paragraph{Does the dataset contain data that might be considered confidential (e.g., data that is protected by legal privilege or by doctor-patient confidentiality, data that includes the content of individuals’ non-public communications)? If so, please provide a description.}
No.

\paragraph{Does the dataset contain data that, if viewed directly, might be offensive, insulting, threatening, or might otherwise cause anxiety? If so, please describe why.}
Unknown.

\paragraph{Does the dataset relate to people? If not, you may skip the remaining
questions in this section.}
Yes.

\paragraph{Does the dataset identify any subpopulations (e.g., by age, gender)? If so, please describe how these subpopulations are identified and
provide a description of their respective distributions within the dataset.}
No.

\paragraph{Is it possible to identify individuals (i.e., one or more natural persons), either directly or indirectly (i.e., in combination with other
data) from the dataset? If so, please describe how}
No.

\paragraph{Does the dataset contain data that might be considered sensitive
in any way (e.g., data that reveals racial or ethnic origins, sexual
orientations, religious beliefs, political opinions or union memberships, or locations; financial or health data; biometric or genetic data; forms of government identification, such as social security numbers; criminal history)? If so, please provide a description.}
No.

\paragraph{Any other comments?}
No.

\subsection{Collection Process}

\paragraph{How was the data associated with each instance acquired? Was
the data directly observable (e.g., raw text, movie ratings), reported by
subjects (e.g., survey responses), or indirectly inferred/derived from other
data (e.g., part-of-speech tags, model-based guesses for age or language)?
If data was reported by subjects or indirectly inferred/derived from other
data, was the data validated/verified? If so, please describe how.}
All data was collected by scraping problems from coding challenge websites, such as Codewars, AtCoder and Kattis.

\paragraph{What mechanisms or procedures were used to collect the data
(e.g., hardware apparatus or sensor, manual human curation, software program, software API)? How were these mechanisms or procedures validated?}
We used off-the-shelf and custom-built scrapers. We manually checked whether scraped data matched text on the websites.

\paragraph{If the dataset is a sample from a larger set, what was the sampling
strategy (e.g., deterministic, probabilistic with specific sampling
probabilities)?}
Some problems we scraped were left out of APPS for various reasons, e.g. they required images to solve, they lacked ground-truth solutions and test cases, or they were duplicate problems.

\paragraph{Who was involved in the data collection process (e.g., students,
crowdworkers, contractors) and how were they compensated (e.g.,
how much were crowdworkers paid)?}
All data was collected by undergraduate and graduate student authors on the paper.

\paragraph{Over what timeframe was the data collected? Does this timeframe
match the creation timeframe of the data associated with the instances
(e.g., recent crawl of old news articles)? If not, please describe the timeframe in which the data associated with the instances was created.}
Data was collected from late 2020 to early 2021 and refined for six months.

\paragraph{Were any ethical review processes conducted (e.g., by an institutional review board)? If so, please provide a description of these review
processes, including the outcomes, as well as a link or other access point
to any supporting documentation}
No.

\paragraph{Does the dataset relate to people? If not, you may skip the remainder
of the questions in this section.}
Yes.

\paragraph{Did you collect the data from the individuals in question directly,
or obtain it via third parties or other sources (e.g., websites)?}
We scraped data via websites where individuals had publicly posted problem solutions.

\paragraph{Were the individuals in question notified about the data collection? If so, please describe (or show with screenshots or other information) how notice was provided, and provide a link or other access point to, or otherwise reproduce, the exact language of the notification itself.}
Users who posted on the Internet were not notified of our collection, because their examples were posted publicly.

\paragraph{Did the individuals in question consent to the collection and use
of their data? If so, please describe (or show with screenshots or other
information) how consent was requested and provided, and provide a
link or other access point to, or otherwise reproduce, the exact language
to which the individuals consented.}
N/A

\paragraph{If consent was obtained, were the consenting individuals provided with a mechanism to revoke their consent in the future or
for certain uses? If so, please provide a description, as well as a link or
other access point to the mechanism (if appropriate).}
N/A

\paragraph{Has an analysis of the potential impact of the dataset and its use
on data subjects (e.g., a data protection impact analysis) been conducted? If so, please provide a description of this analysis, including
the outcomes, as well as a link or other access point to any supporting
documentation.}
No.

\paragraph{Any other comments?}
No.

\subsection{Preprocessing/Cleaning/Labeling}

\paragraph{Was any preprocessing/cleaning/labeling of the data done (e.g.,
discretization or bucketing, tokenization, part-of-speech tagging,
SIFT feature extraction, removal of instances, processing of missing values)? If so, please provide a description. If not, you may skip the
remainder of the questions in this section.}
Yes, as described in Section 3 of the main paper.

\paragraph{Was the “raw” data saved in addition to the preprocessed/cleaned/labeled
data (e.g., to support unanticipated future uses)? If so, please provide a link or other access point to the “raw” data.}
No.

\paragraph{Is the software used to preprocess/clean/label the instances available? If so, please provide a link or other access point.}
Not at this time.

\paragraph{Any other comments?}
No.

\subsection{Uses}
\paragraph{Has the dataset been used for any tasks already? If so, please provide
a description.}
Yes, see the main paper.

\paragraph{Is there a repository that links to any or all papers or systems that
use the dataset? If so, please provide a link or other access point.}
No.

\paragraph{What (other) tasks could the dataset be used for?}
N/A

\paragraph{Is there anything about the composition of the dataset or the way
it was collected and preprocessed/cleaned/labeled that might impact future uses? For example, is there anything that a future user
might need to know to avoid uses that could result in unfair treatment
of individuals or groups (e.g., stereotyping, quality of service issues) or
other undesirable harms (e.g., financial harms, legal risks) If so, please
provide a description. Is there anything a future user could do to mitigate
these undesirable harms?}
We describe how our data collection is legally compliant in Appendix A.

\paragraph{Are there tasks for which the dataset should not be used? If so,
please provide a description.}
N/A

\paragraph{Any other comments?}
No.

\subsection{Distribution}
\paragraph{Will the dataset be distributed to third parties outside of the entity (e.g., company, institution, organization) on behalf of which
the dataset was created? If so, please provide a description.}
Yes, the dataset will be publicly distributed.

\paragraph{How will the dataset will be distributed (e.g., tarball on website,
API, GitHub)? Does the dataset have a digital object identifier (DOI)?}
The dataset is available at \href{https://github.com/hendrycks/apps}{https://github.com/hendrycks/apps}.

\paragraph{When will the dataset be distributed?}
The dataset is currently available.

\paragraph{Will the dataset be distributed under a copyright or other intellectual property (IP) license, and/or under applicable terms of use
(ToU)? If so, please describe this license and/or ToU, and provide a link
or other access point to, or otherwise reproduce, any relevant licensing
terms or ToU, as well as any fees associated with these restrictions.}
The code for our experimental framework is distributed under an MIT license. Where applicable, 

\paragraph{Have any third parties imposed IP-based or other restrictions on
the data associated with the instances? If so, please describe these
restrictions, and provide a link or other access point to, or otherwise
reproduce, any relevant licensing terms, as well as any fees associated
with these restrictions.}
In cases where websites that we scrape data from have copyright policies, we abide by Fair Use according to §107, and we comply with GDPR even though all our problem sources with ground-truth solutions are based in the US. See Appendix A for details.

\paragraph{Do any export controls or other regulatory restrictions apply to
the dataset or to individual instances? If so, please describe these
restrictions, and provide a link or other access point to, or otherwise
reproduce, any supporting documentation.}
No.

\paragraph{Any other comments?}
No.

\subsection{Maintenance}
\paragraph{Who is supporting/hosting/maintaining the dataset?}
Refer to the main document.

\paragraph{How can the owner/curator/manager of the dataset be contacted
(e.g., email address)?}
Refer to the main document.

\paragraph{Is there an erratum? If so, please provide a link or other access point.}
Not at this time.

\paragraph{Will the dataset be updated (e.g., to correct labeling errors, add
new instances, delete instances)? If so, please describe how often, by
whom, and how updates will be communicated to users (e.g., mailing list,
GitHub)?}
We plan to update the dataset with an additional JSON of test cases present in the question text for each problem. This will be available through GitHub.

\paragraph{If the dataset relates to people, are there applicable limits on the
retention of the data associated with the instances (e.g., were individuals in question told that their data would be retained for a
fixed period of time and then deleted)? If so, please describe these
limits and explain how they will be enforced}
No.

\paragraph{Will older versions of the dataset continue to be supported/hosted/maintained?
If so, please describe how. If not, please describe how its obsolescence
will be communicated to users.}
N/A

\paragraph{If others want to extend/augment/build on/contribute to the
dataset, is there a mechanism for them to do so? If so, please
provide a description. Will these contributions be validated/verified?
If so, please describe how. If not, why not? Is there a process for communicating/distributing these contributions to other users? If so, please
provide a description.}
Our dataset could be extended with additional problems that follow the formatting of existing problems.

\paragraph{Any other comments?}
No.

\section{Additional Dataset Information}

\paragraph{Expanded Dataset Comparisons.}
We compared to several datasets in the \citep{NEURIPS2019_7298332f,yu2018spider,Raychev2016ProbabilisticMF,iyer-etal-2018-mapping,Lu2021CodeXGLUEAM} main paper. We continue the comparisons below.
\citet{Ling2016LatentPN} introduce datasets based on Hearthstone and Magic the Gathering card games for code generation. \citet{oda2015ase:pseudogen1} provide a language-to-code dataset using simple code comments. \citet{Zavershynskyi2018NAPSNP} introduce the NAPS dataset for converting pseudocode to code, obtained by crowdsourcing low-level descriptions of programming exercises, and apply machine translation techniques to the problem. Recent anecdotal posts on social media have demonstrated that modern Transformers can in some instances generate JSX code adhering to user requests, but our work provides precision to the discussion through quantitative evaluation. \citet{allamanis2013mining} introduce the GitHub Java Corpus used for performing language modeling on Java code. \citet{raca} do a smaller-scale analysis of code generation but with their limited language-specific training data models ``fail to pass even a single predefined test case'' on their 300 test problems, while with our large training set and test set, trained models can pass tens of thousands of test cases. \citet{Zelle1996LearningTP} and \citet{Tang2001UsingMC} precedes \citet{yu2018spider} by also facilitating the synthesis of database queries, though more recent program synthesis works such as \citet{wang2019rat} use Spider from \citet{yu2018spider}.

\Cref{tab:further_dataset_comparison} compares APPS to Hearthstone \citep{Ling2016LatentPN}, Django \citep{oda2015ase:pseudogen1}, and \citet{Zavershynskyi2018NAPSNP}. `Number of Programs' refers to the number of human-written programs or functions in the dataset, and `Number of Exercises' refers to the number of tasks that the network must solve. These numbers can differ in datasets such as APPS with multiple human-written solutions per exercise.

\paragraph{Excluded Keywords.}\label{appx:excluded_keywords} In creating the GitHub pretraining dataset, we exclude the following keywords to prevent overlap with coding challenge questions similar to those in APPS: `atcoder', `coderbyte', `leetcode', `codeforces', `codewars', `hackerrank', `topcoder', `codechef', `checkio', `HackerEarth', `Programmr', `Exercism', `Codier', `PyBites', `Tynker', `CodinGame', `CodeCombat', `usaco', `IOI', `UVA', `ICFP', `EPIJudge', `SPOJ', `UVaOJ', `judge', `interview', `solution', `coding', `code', `problem', `exercise', `challenge', `algo', `practice', `competitive', `program'.

\section{Additional Results}

\paragraph{Top-5 Performance.} Rather than allowing models to generate just one potential solution, we let models generate five and we choose the best performing solution. Full top-5 performance results are in \Cref{tab:top5full}.

\paragraph{GPT-3.}
We evaluate GPT-3 175B on APPS in a few-shot setting. A separate prompt is used for standard input and call-based questions, and each prompt includes instruction text along with two example questions and solutions from the corresponding question type. The solutions we select for each prompt are the shortest solutions passing all test cases for the corresponding problems. In preliminary experiments, we found that the original GPT-3 175B did not perform well, and the instruct series was necessary for reasonable generations. Inference on the $5,\!000$ problems in the APPS test set is possible for approximately \$500.

We find that GPT-3 only solves $3$ problems out of $5,\!000$: two introductory problems and one interview problem. The two introductory problems are simple interpretation tasks, such as implementing a specified algebraic expression. The interview problem requires higher-level thinking and is show in Figure \ref{fig:interview_question_solved_by_GPT3}. Legitimately solving this problem would suggest nontrivial reasoning. However, it is possible that GPT-3 memorized the solution during pretraining, or that it took a lucky guess based on heuristics in the question.

We also find that GPT-3 tends to handle syntax poorly. In Figure \ref{fig:bad_syntax_GPT3}, we see a generation with improper formatting of an if-else block, which causes a syntax error. By contrast, GPT-2 0.1B has fewer syntax errors with three orders of magnitude fewer parameters, and GPT-Neo 2.7B has almost no syntax errors. These two models are fine-tuned on GitHub and the APPS training set.

\begin{figure*}
    \textbf{Problem} \\
    Mahmoud and Ehab play a game called the even-odd game. Ehab chooses his favorite integer $n$ and then they take turns, starting from Mahmoud. In each player's turn, he has to choose an integer $a$ and subtract it from $n$ such that:  $1 \leq a \leq n$.  If it's Mahmoud's turn, $a$ has to be even, but if it's Ehab's turn, $a$ has to be odd. 

    If the current player can't choose any number satisfying the conditions, he loses. Can you determine the winner if they both play optimally?
    
    \vspace{2mm}
    -----Input-----
    
    The only line contains an integer $n$ $(1 \leq n \leq 10^9)$, the number at the beginning of the game.
    
    \vspace{2mm}
    -----Output-----
    
    Output ``Mahmoud'' (without quotes) if Mahmoud wins and ``Ehab'' (without quotes) otherwise.
 \\
 \\
 \textbf{Model Output}
 \begin{small}
 \begin{minted}{python}
n = int(input())
if n %
    print('Mahmoud')
else:
    print('Ehab')
\end{minted}
 \end{small}
    \caption{The sole interview problem solved by GPT-3. Legitimately solving this problem requires nontrivial reasoning about the proposed game's branching possibilities.}
    \label{fig:interview_question_solved_by_GPT3}
\end{figure*}

\begin{figure*}
 \begin{small}
 \begin{minted}{python}
a, b = int(input()), int(input())
if a == b:

print(''.join(map(str, [a, b])))

else:

print(''.join(map(str, [a, b])))
\end{minted}
 \end{small}
    \caption{Oftentimes, GPT-3 generates outputs with incorrect syntax, such as the above code. Even though the code is reasonable, bad formatting of the if-else block causes a syntax error.}
    \label{fig:bad_syntax_GPT3}
\end{figure*}

\paragraph{More Generated Code Examples.} Additional examples of generated code can be found in \Cref{fig:supp_sample_1}, \Cref{fig:supp_sample_2}, and \Cref{fig:supp_sample_3}.

\begin{figure*}
    \textbf{Problem} \\
    Given is a directed graph G with N vertices and M edges. The vertices are numbered 1 to N, and the i-th edge is directed from Vertex A\_i to Vertex B\_i. It is guaranteed that the graph contains no self-loops or multiple edges. Determine whether there exists an induced subgraph (see Notes) of G such that the in-degree and out-degree of every vertex are both 1. If the answer is yes, show one such subgraph. Here the null graph is not considered as a subgraph.
    \\
     \\
 \textbf{Model Output}
 \begin{small}
 \begin{minted}{python}
import sys

sys.setrecursionlimit(10 ** 7)
input = sys.stdin.readline
f_inf = float('inf')
mod = 10 ** 9 + 7

def resolve():
    n, m = map(int, input().split())
    edge = [[] for _ in range(n)]
    for _ in range(m):
        a, b = map(int, input().split())
        edge[a - 1].append(b - 1)
        edge[b - 1].append(a - 1)

    def dfs(u):
        stack = [u]
        seen = [False] * n
        while stack:
            u = stack.pop()
            if seen[u]:
                continue
            seen[u] = True
            for v in edge[u]:
                if seen[v]:
                    continue
                seen[v] = True
                stack.append(v)

    res = None
    for i in range(n):
        if not seen[i]:
            seen[i] = True
            res = dfs(i)
    if not res:
        print(-1)
    else:
        print(res)

if __name__ == '__main__':
    resolve()
\end{minted}
\end{small}
    \caption{An example generated by GPT-2 1.5B. The model can sometimes compose function calls.}
    \label{fig:supp_sample_1}
\end{figure*}

\begin{figure*}
    \textbf{Problem} \\
    Given is a tree G with N vertices. The vertices are numbered 1 through N, and the i-th edge connects Vertex a\_i and Vertex b\_i. Consider painting the edges in G with some number of colors. We want to paint them so that, for each vertex, the colors of the edges incident to that vertex are all different. Among the colorings satisfying the condition above, construct one that uses the minimum number of colors.
 \\ 
 \\
 \textbf{Model Output}
 \begin{small}
 \begin{minted}{python}
import sys
from collections import deque, defaultdict, Counter
from itertools import accumulate, product, permutations, combinations
from operator import itemgetter
from bisect import bisect_left, bisect_right
from heapq import heappop, heappush
from math import ceil, floor, sqrt, gcd, inf
from copy import deepcopy
import numpy as np
import scipy as sp

INF = inf
MOD = 1000000007

n = int(input())
ab = [list(map(int, input().split())) for i in range(n - 1)]

graph = [[] for i in range(n + 1)]
for a, b in ab:
    graph[a].append(b)
    graph[b].append(a)

visited = [False] * (n + 1)
color = [0] * (n + 1)
color[0] = 1
stack = [1]
while stack:
    x = stack.pop()
    for y in graph[x]:
        if visited[y]:
            continue
        visited[y] = True
        color[y] = color[x] + 1
        stack.append(y)

ans = min(color)
print(*ans, sep="\n")
\end{minted}
\end{small}
    \caption{An additional generated example. Here, the model output has many import statements.}
    \label{fig:supp_sample_2}
\end{figure*}

\begin{figure*}
    \textbf{Problem} \\
    There is a special square room with mirrors on each of the four walls.  Except for the southwest corner, there are receptors on each of the remaining corners, numbered 0, 1, and 2. The square room has walls of length p, and a laser ray from the southwest corner first meets the east wall at a distance q from the 0th receptor. Return the number of the receptor that the ray meets first.  (It is guaranteed that the ray will meet a receptor eventually.)
    \\
    
    Example 1:
    
    Input: p = 2, q = 1
    
    Output: 2
    
    Explanation: The ray meets receptor 2 the first time it gets reflected back to the left wall.
    \\

    \textbf{Starter Code}
    
    \begin{small}
\begin{minted}{python}
class Solution:
    def mirrorReflection(self, p: int, q: int) -> int:
\end{minted}
\end{small}

 \textbf{Model Output}
 \begin{small}
\begin{minted}{python}
class Solution:
    def mirrorReflection(self, p: int, q: int) -> int:
        self.p = p
        self.q = q
        self.mirrors = [(0, 1), (0, -1), (1, 0), (-1, 0)]
        return self.mirrors[self.p - 1][self.q - 1]
\end{minted}
\end{small}
    \caption{An example using the call-based format with some starter code.}
    \label{fig:supp_sample_3}
\end{figure*}